\documentstyle[12pt,aaspp4]{article}

\lefthead{Clemens, et al}
\righthead{Lower Main Sequence}
\begin {document}
\title{The Lower Main Sequence and the Orbital Period Distribution
of Cataclysmic Variable Stars}

\author {J. Christopher Clemens}
\author {I. Neill Reid
\footnote
{ Visiting Research Associate, Observatories of the Carnegie Institution of
Washington}}
\author {John E. Gizis}
\affil {Palomar Observatory, 105-24, California Institute of Technology, Pasadena, CA 91125,
e-mail: jcc@astro.caltech.edu, inr@astro.caltech.edu, jeg@astro.caltech.edu}
\author {M. Sean O'Brien}
\affil {Department of Physics and Astronomy, Iowa State University, Ames, IA 50011, e-mail:msobrien@iastate.edu}
\begin {abstract}
   
The color-magnitude diagram of the lower main sequence, 
as measured from a volume-limited sample of nearby stars,
shows an abrupt downward jump between $M_V \sim 12$ and $13$.  
This jump indicates that the observed mass-radius relationship 
steepens between 0.3 and 0.2 $M_\odot$, but theoretical
models show no such effect.
It is difficult to isolate the source of this disagreement:
the observational mass-radius relationship relies upon 
transformations that may not be sufficiently accurate, while the
theoretical relationship relies upon stellar models 
that may not be sufficiently complete, particularly in their
treatment of the complex physics governing the interior equation-of-state.

    If the features in the observationally 
derived mass-radius relationship 
are real, their existence provides 
a natural explanation for the well-known gap 
in the orbital period distribution of cataclysmic variables.
This explanation relies only upon the observed mass-radius relationship of
low-mass stars, and does not require {\sl ad hoc} changes in magnetic
braking or in the structure of cataclysmic variable secondaries. 
If correct, it will allow broader application of cataclysmic 
variable observations to problems of basic stellar physics.

\end {abstract}


\section {Introduction}

   As physics laboratories, stars leave something to be desired.  The
number of observable stellar parameters is usually far outmatched
by the number of unknowns contained in a stellar model.  To progress, 
theorists must eliminate variables through the use of relations 
based on the best possible constitutive physics.  When, inevitably, the 
resulting models fail to match all the observational details, it is 
difficult to know which physics to adjust. This is unfortunate, because
it is precisely in the mismatch between theory and observation that 
the opportunity to improve physical insight lies.

   The best solution to this dilemma is to increase the number and variety
of observational constraints.   Many interesting experiments, not
of our design, make this possible.  For stellar astronomers, the most dramatic, 
and potentially revealing, are those offered by interacting binary stars,
particularly the cataclysmic variable stars.
These are binaries consisting of a low-mass secondary (typically $< 1 \rm M_\odot$) 
transferring matter, usually through an accretion disk, onto a white dwarf 
star.  Warner (1995; see also \cite{P84}), has written an excellent and comprehensive
review of these systems. 

   The secondaries in cataclysmic variables, which  are being slowly stripped
of mass, offer a perspective on the lower main sequence not available from
single low-mass stars, which spend their entire lives at
more-or-less constant mass.  Unfortunately, the complexities of the mass
transfer introduce a spectacular, often bewildering array of observational 
properties that must be interpreted with caution.  These include 
complex geometric modulations, rapid photometric flickering,
and outbursts with amplitudes up to 19 magnitudes. 
Fortunately, it is usually possible to measure at least one parameter,
the orbital period, with high accuracy.

In this paper, we will explore the connection between 
cataclysmic variable orbital periods and observations of low-mass
single stars, under the
simplest of assumptions.  In particular, we will examine
a downward jump that appears around $M_V \sim 12$ in color-magnitude diagrams 
of population I stars. The location of this jump, and 
its effect on the inferred mass-radius relationship, 
provide a reasonable and natural explanation for the most distinctive
feature of the cataclysmic variable orbital period distribution, the 
period gap.

\section{The Lower Main Sequence}

   In this section, we will examine the morphology of the lower
main sequence using color-magnitude diagrams of a volume-limited sample 
of nearby stars defined by Reid \& Gizis (1997).  
The ultimate purpose of this exploration is to show that the observations
contain evidence for changes in the slope of the mass-radius 
relationship---changes not reproduced 
by theoretical models of low-mass stars.  
To construct the mass-radius diagram,
we will transform the observations into the theoretical plane 
using the best available bolometric corrections and effective temperature
scale (\cite{L96}).  Then we will investigate the implied shape of 
the mass-radius relationship using the empirical mass-magnitude relationships
of Henry \& McCarthy (1993).  The most uncertain component of this 
process, and the only one which depends explicitly upon theoretical models, is
the color-temperature transformation.  Consequently, we will discuss in 
detail whether the discrepancy between theory and observation results 
from this problematic transformation, or whether it arises from 
inadequacies in the stellar models.

\subsection{The Observations}

  Reid, Hawley \& Gizis (1995) have recently completed a spectroscopic survey 
of M-dwarfs drawn from the most recent version of the Catalogue of Nearby 
Stars (\cite{GJ91}).  In the course of this survey,
Hawley, Gizis \& Reid (1996) noted that the color-magnitude diagram of the
stars with the best trigonometric parallaxes and photometric measurements
shows a step or break  near $(V-I) \sim 2.9$.  This is apparently
the same feature noted by Leggett, Harris, \& Dahn (1994) in their observations of 
Hyades dwarfs.  It also appears in published color-magnitude diagrams
of NGC 2420 and NGC 2477, as  measured with the {\sl Hubble Space Telescope} by
von Hippel et al. (1996).

  Subsequent to the M-dwarf survey, Reid \& Gizis (1997) showed that the 
survey stars north of $\delta = -30^o$ comprise a statistically
complete sample to 10 parsecs for $M_V \le +14$.  Then they used the 
nearest 8 parsecs of the sample to study the fraction of 
multiple stars and its effect on the 
stellar luminosity function.  In this paper we will use the
same 8 parsec sample to examine the mass-radius relationship of
stars on the lower main sequence.  The details of this sample, 
along with a complete
listing of the stars and their properties, can be found in 
Reid \& Gizis (1997).  

  Our investigation does not strictly require a volume-limited
sample, but the 8 parsec sample of Reid \& Gizis (1997) offers two
virtues: it was not selected on any subjective basis which might bias the
shape of the color-magnitude diagram, and its ($M_V,(V-I)$)
main sequence was fitted with a composite relation 
by Reid \& Gizis (1997) before any of us
recognized the connection to cataclysmic variables.  We have
used this fit, rather than calculating our own, so that there is no
chance for bias to enter in the selection of fitting points.

  The upper panels of Figure 1 show 
the ($M_V,(V-I)$) and ($M_K,(I-K)$) main sequences of
stars from the 8 parsec sample.  The photometry and distance
determinations are compiled from
a variety of published sources, as tabulated by Reid \& Gizis (1997).
The open circles denote stars for which the photometry
is of lower quality, either because it was transformed from another 
photometric system into the Kron-Cousins system, or because the stars
are components of small separation ($<5$'') binaries. 
For the solid points, Reid \& Gizis (1997) note that the photometric
accuracy in $(V-I)$ is 
0.02 magnitudes, but the observed dispersion is 0.09, implying some 
intrinsic scatter in their properties.  
The average dispersion in $M_V$ is $\sim 0.35$ magnitudes, but changes
with the slope of the main sequence.

  In the bottom panel of Figure 1, we have augmented the data with 
distances from Hipparcos (ESA 1997)
to better define the main sequence
above $M_V \sim 11$.  This diagram shows intriguing suggestions of
structure in the main sequence near $M_V \sim 8.5$,
but unfortunately Hipparcos does not reach faint enough to
help with the region of primary concern in this paper.
Consequently, we have not used the Hipparcos data in any of 
the analysis which follows.

  The step-like feature discussed by Hawley et al. (1996)
appears in all three of the panels in Figure 1.
In the ($M_V, (V-I)$) plot, we have added a dotted line representing
the composite fit to the data by Reid \& Gizis (1997).  
This fit maintains a roughly constant
dispersion in $(V-I)$, and is described by the following formulae:

\begin{equation}
M_V = \left\{ \begin{array}{lr}
1.087 + 6.226 (V-I) - 1.340 (V-I)^2 + 0.187 (V-I)^3 
& {\rm for}~(V-I) < 2.635 \\
& (\sigma_V = 0.35~\rm{mag}) \\
\\
5.48 (V-I) - 2.83 
& {\rm for}~2.635 \le (V-I) < 3.00 \\
& (\sigma_V = 0.46~\rm{mag}) \\
\\
6.07  + 2.441 (V-I) + 0.0233 (V-I)^2 
& {\rm for}~(V-I) \ge 3.00 \\
& (\sigma_V = 0.26~\rm{mag})
\end{array}\right.
\end{equation}

\subsection{Comparison to Theoretical Models}

  In order to compare the observations in Figure 1 to theoretical models, 
we have translated the data from color and magnitude
to luminosity and temperature.  
The transformations we applied are based on the bolometric corrections
and temperature scale derived by Leggett et al. (1996), who 
used flux-calibrated spectra and trigonometric parallaxes
to measure the bolometric correction for 16 M-dwarfs 
with colors in the range $1.56 < (V-I) < 4.26$. 

  To apply their measurements to the 8 parsec sample, we have
fitted a parabola to their values of bolometric corrections and $(V-I)$ 
colors of the 
stars they identify as disk population.
We find 
\begin{equation}
BC_V \qquad = \qquad 0.477 \quad - \quad 0.765(V-I) \quad - \quad
0.099(V-I)^2, 
\end{equation}
where $BC_V$ is the Johnson V band bolometric correction. 
We applied bolometric corrections calculated from 
this equation to the absolute V magnitudes of our sample and 
converted the result to ${\rm log}(L / L_\odot)$  using $M_{bol, \odot} = +4.75$.

  Leggett et al. (1996) measured temperatures for their sample
by fitting the observed spectra with improved versions of the
atmospheric models of Allard \& Hauschildt (1995).
Again we fitted a parabola to their temperature measurements of disk stars 
to get the color-temperature relation:
\begin{equation}
T_{eff} \qquad = \qquad 5757.8 \quad - \quad 1218.6(V-I) \quad + \quad
116.9(V-I)^2.
\end{equation}
In this instance we weighted our fit according to the
published error estimates for the temperature values.
   
  Our fitting formulae reproduce the measurements of Leggett et al. (1996)
with a dispersion of 41 K in the temperature and 0.06 magnitudes in
the bolometric correction.  Leggett et al. estimate the 
errors in their temperatures to be 150-250 K. 

  The use of parabolic fits in both cases introduces implicit assumptions
about the smoothness of the transformations, which are sparsely sampled by the
Leggett et al. (1996) data.  Considering the smoothness of similar, but better sampled
transformations (Bessell 1995; Dahn et al. 1992) and of theoretically
calculated ones (Allard \& Hauschildt 1995; Brett 1995) these 
are probably safe assumptions, but we will return to 
this issue in Section 2.4.1.

  In Figure 2, we plot the result of our transformations,
along with a variety of stellar models.  For clarity we have 
divided the models between 2 plots, both of which show the same 
observations. We have estimated the errors in temperature
by propagating the dispersion in $V-I$, which overestimates the measurement
error, through equation 3, and adding the dispersion in our
temperature fit (in quadrature).  We get 62-77 K, with the higher value 
applying to larger temperatures, where the ($T_{eff},(V-I)$)
relation is steeper.  We estimated the luminosity errors in a similar
manner using equation 2, but also included the dispersion in $M_V$, which completely 
dominates the other terms.  The error in ${\rm log}(L/L_\odot)$ ranges from
0.12 to 0.19, reflecting the changes in the $\sigma_V$ of equation 1.
For reference, both plots also show the 
composite relation of Reid \& Gizis (1997)  transformed in the
same manner as the data.

 It is encouraging that all of the models show a downturn more-or-less
resembling the feature in the data.  
The atmosphere and envelope
calculations in this temperature domain are notoriously 
difficult (Allard \& Hauschildt 1995; Brett, 1995), and must incorporate
the complex effects of molecule formation upon opacities (Alexander \& Ferguson 1994)
and the envelope equation-of-state (Saumon, Chabrier, \& Van Horn 1995).  
It is a triumph that the models, particularly those of Baraffe et al. (1995) and 
Baraffe \& Chabrier (1996) match the data
so well.  Nonetheless, the downward step in the 
data is more abrupt than that shown by any of the models.
In the following section, we investigate the effect of this
abruptness on the mass-radius relationship.

\subsection{Implications for the Mass-Radius Relationship}

  The shape of the observed color-magnitude diagrams in Figures
1 and 2 implies that a rapid drop in stellar radius occurs between
$M_V \sim 12$ and $13$.  To see whether it also implies a steepening
of the mass-radius relationship, we have translated the   
data to the mass-radius plane using the same method as Hoxie (1973).
First we used
\begin{equation}
R = ({L \over 4 \pi \sigma})^{1/2}T^{-2},
\end{equation}
to calculate  radii directly for the temperature and luminosities
plotted in Figure 2.
Then, to get masses, we used the 
empirical (mass, $M_V$)  relation of Henry \& McCarthy (1993), which 
is based on measurements for 37 stars in visual binaries.  We chose the 
Henry \& McCarthy relation because it is empirical, but the theoretical 
relations of  Kroupa, Tout, \& Gilmore (1993) and 
of Chabrier, Baraffe, \& Plez (1996)  yield similar 
results (shown in Figure 9).

Figure 3 shows the resulting mass-radius relationship.  The lines and 
symbols are the same as in Figure 2, except here we show only the
models of Baraffe \& Chabrier (1996).  We have placed 
representative error bars on the transformed fit of Reid \& Gizis (1997).  
In the radius direction
they represent the temperature and luminosity errors (described earlier)
propagated through the radius calculation.  In spite of the sensitivity of 
radius to temperature, the radius error
is dominated by the scatter in $M_V$.
Consequently, the error
bars shown do not differ substantially from the scatter 
in the individual mass-radius points.  Even if we increase the
temperature error to 150 K, the estimate Leggett et al. (1996)
give for most of their individual measurements, it does not 
significantly increase the radius error we calculate.
The mass error bars are the quadrature sum of the dispersion in 
the 8 parsec data $M_V$ measurements
and Henry \& McCarthy's (1993)
published dispersions for the (mass, $M_V$) relation. The latter are
the largest contributors to the resultant error in mass.

Figure 3 illustrates the main point of this section:
the mass-radius relationship of stars in the 8 parsec
sample shows evidence for changes in slope not present in the 
theoretical models.  The changes appear artificially 
abrupt in the dotted-line fit of Figure 3, reflecting 
the abrupt changes in the composite relations, both for
($M_V, (V-I)$), and for (mass, M$_v$). Otherwise, Figure 3
displays a realistic depiction of the empirical mass-radius relation 
of the 8 parsec sample, inclusive of any systematic 
errors in the photometry and 
transformations. We discuss the possible effects of these
in the next section.

\subsection{Discussion}  

The differences between the theoretical models and the transformed 
observations in Figure 3
are not large, but we will show in section 3 that
their effect on the orbital period evolution
of cataclysmic variables is profound.  Consequently, it is important to
establish how much faith ought to be placed in the observations 
and how much in the models.  First, we will consider whether the slope
changes in the observed mass-radius relationship might
be generated as artifacts either from the photometric data
or the transformations we have applied.  
We will not be able to completely exclude this possibility.  Second, 
we will investigate the models and 
find that the physical parameters the modelers vary have substantial
effects on the way model temperatures change with mass, but
very little effect on the way model radii change with mass.
As a result, the possibility of changes in the slope of 
the mass-radius relationship is not explicitly 
addressed by current models, and therefore 
cannot be ruled out.  

\subsubsection{The Observations and Transformations}

  Figure 4 illustrates two partially independent checks
on the reality of the features in the mass-radius relationship
of the 8 parsec sample.  First, we have plotted the masses and radii
of binaries tabulated by Popper (1980).  With the exception of
the eclipsing binaries 
CM Dra and YY Gem, the Popper data do not represent fundamental radius 
measurements. The radii were calculated via the same method we have used,
but with different bolometric corrections and temperature transformations.
The Popper data are consistent with the shape we have measured
for the mass-radius relation, but
contain too few fundamental points to confirm the
changes in slope.  Furthermore, the most useful eclipsing system, CM Dra,
is probably metal poor (Metcalfe et al. 1996; Gizis 1997),
and inappropriate for comparison to our population I data.

  For the second test illustrated in Figure 4, we have 
repeated the entire set of calculations required to generate our
mass-radius relationship in a different photometric 
bandpass.  This yields a separate measure of the observed mass-radius 
relationship, with independent photometric colors. First, we 
fitted the (BC$_k$, (I-K)) and (T$_{eff}$, (I-K)) data of Leggett 
and used these fits to transform the (M$_k$, (I-K)) data shown in 
the upper right panel of Figure 1.  From this we constructed
a mass-radius relation, this time using the (mass, M$_k$)
relations of Henry \& McCarthy (1993).  
This exercise does not address the question of systematic errors in the
masses, temperatures or distances,  which are common to both sets of 
transformations.  The circles in Figure 4
show the final result. For reference, the dotted line is the same as in 
Figure 3. While the K-band mass-radius relation shows the same morphology as in 
the V-band, the slope changes are more muted, and the deviation from 
theoretical models smaller. These differences give us a feel for
the size of systematic differences between the V and K band photometry and
transformations, but they do not add any weight to the observed features 
in the mass-radius relationship.

  In addition to the reality checks depicted in Figure 4, we have directly
explored 
the possibility that errors in the transformations could generate artifacts
in the mass-radius relation.  For instance, the parameterized 
(mass, $M_V$) relation of Henry \& McCarthy contains two inflection
points as does the parameterized color-magnitude diagram from Reid and Gizis. 
We have tried moving these inflections points artificially
to see if they can cancel to yield a mass-radius relation more like that
of the theoretical models.  This does not work very well: we 
must either move the inflection points so far that the fit to the 
slope of the (mass, $M_V$) data is unacceptably poor, or we must
introduce a discontinuity in the mass-radius relation near 0.4 R$_\odot$.
Furthermore, ten of the stars included in our mass-radius relation are 
from the Henry \& McCarthy sample, and they show evidence for slope 
changes without recourse to the parameterized (mass, $M_V$) relation. 
Consequently it is unlikely that the observed slope changes are the result of
systematic errors in the (mass, $M_V$) relation alone.

  We have also experimented with the color-temperature transformation, 
which is the only transformation we have used that relies on theoretical 
models. 
We mentioned earlier that the shape of our transformations is not well sampled
by the data of Leggett et al. (1996).   
Consequently we have experimented with changes required to transform the 
color-magnitude diagram into exact correspondence with the model of 
Baraffe \& Chabrier (1996).  Given the errors Leggett et al. (1996) quote for
the temperatures, their data could easily accommodate a transformation which 
contains changes in slope that exactly compensate for the
temperature differences between the model and data.  
Thus we cannot formally rule out the possibility that unsampled 
features in our temperature scale have produced artifacts in the mass-radius 
relationship.

However, when we substitute the better-sampled temperature transformation of 
Bessell (1995), the changes in the mass-radius
relation remain.  To straighten the
mass-radius relation, we would need a color-temperature transformation
like that plotted in Figure 5.  For reference, Figure 5 also shows
the Bessell (1995) transformation and our fit to the Leggett et al. (1996).
data.  Obviously, the features required to 
bring the data into correspondence with the models would be difficult to hide 
within the Bessell transformation, but not impossible, especially 
if the atmospheric models
used to measure the temperatures introduce additional errors.  
Nonetheless, neither the models nor the measurements
offer any evidence for the existence of abrupt features in the
color-temperature transformation; to suppress the changes in the 
mass-radius relation we would have to invoke them arbitrarily.

\subsubsection{The Models}

Finally, we will consider 
whether the differences
between the theoretical and observed mass-radius relationship
can be accommodated by uncertainties in the theoretical
models. In Figure 6, we plot mass-radius
relations for a wider variety of models. Like the mass-radius relation 
of Baraffe \& Chabrier (1996), they are all relatively straight; none show the
features we observe in the 8 parsec sample.
It is possible 
that their straight appearance is due to a fundamental and unalterable 
property of low-mass stars.  
If this is true, and can be demonstrated, it will be 
very useful. It will provide valuable constraints
on the transformations we have used,
and might significantly improve our knowledge of
the (mass, $M_V$) relation, and consequently the mass function of the lower
main sequence.

More likely, the similarity of the mass-radius relations are
due to common simplifications employed by the models that 
produce them.  
For low-mass fully-convective and almost fully-convective model stars,
the radiative region at the surface establishes the thermal equilibrium 
configuration, and thereby the nuclear energy generation rate
(see Cox \& Giuli 1968).  
Hence the luminosities of these models are extremely sensitive to the surface
boundary conditions. However, changes in the surface boundary conditions
affect the luminosity almost exclusively through changes in the 
effective temperature, rather than in radius.  

Dorman, Nelson \& Chau (1989)
illustrate this last point 
by considering the effect of
modifying the surface opacities.  An increase in opacity results in 
a decrease in the surface temperature, because the optical depth rises more 
steeply with pressure ($dP/d\tau \sim g/\kappa$) and our line of 
sight cannot penetrate to as
high a temperature as it did before.  This drop in temperature will
lower the luminosity of the model star unless it expands to compensate.  
But even a small expansion will radically change the nuclear reaction 
rates in the core, because of their sensitive dependence on the central 
temperature.
Consequently, the net effect of the increased opacity is an 
appreciably lower effective temperature and luminosity, but   
only a slightly larger radius.
It is for this reason that model subdwarfs have higher temperatures and
luminosities, but essentially the same radii as model dwarfs of equivalent 
mass (see Schwarzschild 1958).

This insensitivity of model radius to surface physics is not
confined to opacities alone.  
The steepening of the main sequence 
between $-1.8 < {\rm log}(L/L_\odot) <  -1.5$ exhibited
by all of the models shown in Figure 2 was first calculated by 
Copeland, Jensen, \& Jorgensen (1970)
by including the effects of H$_2$ molecules on the 
envelope equation-of-state. The presence of an H$_2$ dissociation region
near the model surface decreases the adiabatic gradient, making the 
star appear hotter and more luminous than it would otherwise.
Figure 3 of Copeland et al. (1970) shows that their models including 
this effect are displaced along lines of almost constant radius from those
which neglect it.

These examples illustrate the weak dependence of radius on 
changes in the surface physics.
For a low-mass, fully-convective model star, the radius is sensitive
primarily to
the interior composition, equation-of-state,  and the nuclear reaction rates.  
We can see this more clearly by considering the classic
study of Hayashi \& Hoshi (1961),
and Hayashi (1961).  They showed that for fully convective, and therefore
isentropic, models, the quantity,
\begin{equation}
E=4\pi({\mu \over N_Ak})^{5/2}G^{3/2}KM^{1/2}R^{3/2}, 
\end{equation}
has a constant value of 45.48.
In this equation, $\mu$ is the mean molecular weight, $N_A$ is 
Avogadro's number, $k$ is Boltzmann's constant, G is the gravitational
constant, and K is the
coefficient of the polytropic law which represents these stars:
\begin{equation}
P=KT^{5/2}.
\end{equation}
Models of fully-convective pre-main sequence stars of fixed
$M$ shrink to smaller 
radius at roughly constant temperature until they reach
an equilibrium between luminosity and nuclear generation rate.
Thus the final radius depends on the internal structure.
An alternative way to see this is to 
recognize that $K$ and $\mu$ in equation 5 depend primarily
on interior properties, and $R$ must adjust to their values such
that $E = 45.48$.

While published models have explored carefully the changes in the surface
physics for models of different mass, they have not explored the limits allowed 
by changes in interior structure or physics that might occur with mass.  
These are possible if, for example, 
stars develop gradients in $\mu$ due to diffusion or nuclear burning. 
\footnote{ Similar changes
might result if the nuclear reaction 
rates change radically in their dependence on temperature or density--- a 
possibility we mention only for completeness.}
In this case we might expect
the fully convective stars to be homogeneous, and the higher mass stars
with radiative cores to sustain a $\mu$ gradient. The boundary between 
models with convective cores and models that are fully convective 
is 0.3 M$_\odot$, coincident with the change in slope of the mass-radius 
relation of Figure 3. If models do not properly account for 
changes in interior composition above and below this mass, it might
help explain the discrepancy with the observations.

Another possibility is that the non-ideal corrections to the interior
equation-of-state, which become large at low masses, might change the
slope of the mass-radius relation.  An extreme example of non-ideal
effects was suggested for the sun by 
Pollock \& Alder (1978) who, in an attempt to resolve the
solar neutrino problem,  proposed that  
iron might be immiscible with hydrogen in the solar
interior.  Later calculations by Alder, Pollock, \& Hansen (1980) and 
Iyetomi \& Ichimaru (1986), 
showed that the temperature in the solar core
is about three times hotter than the critical point
in their phase diagrams, so separation should not occur there.  
However, comparison of their phase diagrams for an 
iron-hydrogen plasma
to the run of temperature and pressure in low-mass stellar models
plotted by Dorman et al. (1989) shows that the conditions
they calculate for phase separation are present near the center of
stars between 0.1 and 0.2 $M_\odot$.  

The point of this discussion is not to suggest that any one of these examples 
is responsible for the discrepancy between the models and observations,
but rather to suggest 
that the limits to our knowledge inevitably translate into 
uncertainties in the radii of our stellar models. It is
important to know the size of these uncertainties,  
because even small changes in the mass-radius relationship can
significantly affect the orbital period evolution of cataclysmic
variables.

\section{The Cataclysmic Variable Orbital Period Distribution}

  The downward jump in the main sequence we have presented,
and the change in the mass-radius relation it implies, bear an obvious 
relevance to the orbital period evolution of cataclysmic variables.
In order to maintain the mass transfer which characterizes the
cataclysmic variables, the secondaries in these binaries must 
fill their Roche lobes.  This condition immediately establishes a relation
between the mean density of the secondaries and the orbital period 
(Faulkner, Flannery and Warner
1972; Eggleton 1983)--- a relation  which is 
essentially independent
of the mass of the accreting white dwarf (as long as it is the more
massive component).
Consequently, the mass-radius
relationship of the secondary
uniquely fixes the orbital period
of a cataclysmic binary.
As the binary evolves, and the secondary loses mass,
changes in the slope of the mass-radius relationship, inasmuch as they affect 
the way the secondary responds to mass loss, will change the way 
in which the orbital period evolves.

  Theoretical investigations of cataclysmic variable evolution 
(Paczynski 1967; Faulkner 1971; Vila, 1971; Rappaport, Joss, \& Webbink 1982; 
McDermott \& Taam 1989; Hameury 1991, and many more) generally 
rely upon theoretical models
of the secondary star to provide the mass-radius relationship.
As we have seen, the theoretical mass-radius relationships are
all essentially featureless between 0.6 and 0.1 M$_\odot$.
In contrast to this approach, 
we will explore the effects of the observationally derived
mass-radius relationship upon the 
evolution of cataclysmic variables.  
We will show that
the mass-radius relation we have constructed implies 
that cataclysmic variables 
in the orbital period range 3.5 to 2.0 hours must either evolve
more quickly, or transfer mass more slowly, than binaries above and below 
this range.   
This result derives from the steeper slope of the observed mass-radius
relation between 0.3 and 0.16 M$_\odot$; a small reduction in mass
results in a larger shrinkage in radius, and therefore a larger reduction
in orbital period.

 The critical assumption in this approach is that the mass-radius
relation measured for single stars can be applied to the mass-losing
secondaries in cataclysmic variables. This is equivalent to assuming
that the secondary remains in thermal equilibrium in spite of
the effects of mass-loss and irradiation. There are sound reasons,
both observational and theoretical, to suppose that this is true
for at least some of the cataclysmic variables.  We will 
discuss these reasons before presenting the results of
our study. 

Beyond the assumption of thermal equilibrium, the results of this section 
depend only upon the observed mass-radius relationship, the size of 
Roche equipotentials,  and Kepler's third law.  They do not depend upon
theoretical models of low-mass stars, nor upon the theoretical
discussions of the previous section.  
The ultimate promise of the investigation we present is to establish a 
better observational link between low-mass stars and cataclysmic variables,
thereby increasing the number of useful observational constraints on 
low-mass stellar models.

\subsection{The Assumption of Thermal Equilibrium}

Observationally, there is evidence that at least some, and maybe most,
cataclysmic variables secondaries have radii like normal main sequence
stars (Webbink 1990; Ritter 1980; Marsh 1990; Wade \& Horne 1988).  
Webbink (1990) constructed an empirical mass-radius relationship for 24 
eclipsing cataclysmic binaries and found it to be a good match to the
theoretical models.  He has kindly provided his results in tabular form 
(Webbink 1997) for comparison to our mass-radius relationship. 
As Figure 7 shows, his measurements are in excellent agreement
with the mass-radius relationship of the 8 parsec sample.  Clearly the
effects of mass loss
have not drastically increased the radii of the secondaries in these
systems, but the data do not exclude small effects that might
have significant consequences for our results.  Consequently, we will
also discuss theoretical issues concerning the assumption of thermal 
equilibrium. 

Intuitively, we expect that mass loss from the secondary will
cause it to have a radius larger than a main sequence star of equivalent
mass if the mass is removed faster than the star can adjust its 
equilibrium structure.  Hence, when the mass loss timescale, defined as 
$M_2/\dot{M}_2$, drops below the Kelvin-Helmholtz timescale, the
assumption of thermal equilibrium may become invalid.  For a secondary
in a cataclysmic variable with an orbital period of 3 hours,
the Kelvin-Helmholtz timescale is $2.4 \times 10^8$ yr, which yields 
a limiting mass loss rate of $1.3 \times 10^{-9} M_\odot/{\rm yr}$.
For rates higher than this, the main sequence mass-radius relation may
no longer apply.

Warner (1995) presents a calibration of average mass transfer rates
in cataclysmic variables versus the absolute magnitudes of their disks
(his Figure 9.8, see also Smak 1989,1994).
This relationship suggests that the dwarf novae have average mass 
transfer rates near or below this limit.  Thus we may expect our application
of the mass-radius relation for single stars to apply to some, if not
all, cataclysmic variables.  This conclusion is dependent upon the degree to
which it is valid to infer mass transfer rates from the brightness of 
accretion disks (Smak 1994).  It also assumes that there have been no
recent episodes of much higher mass-transfer that might have temporarily
driven the the secondary out of thermal equilibrium.

Another effect that might cause the secondary to have a radius larger
than a single star of equivalent mass is irradiation.
The effects of irradiation by the companion and disks are similar to other
changes in the surface boundary condition of stellar models
(Tout et al. 1989).  As we discussed in section 2.4, the surface 
boundary condition affects primarily the temperature of the models, not 
the radius.  Thus we intuitively expect irradiation to have large 
effects on the temperature of secondaries in cataclysmic variables, 
but little effect on radii.  This intuition is borne out by detailed 
models (Podsiadlowsky 1991;Hameury et al. 1993) up until the
incident flux is roughly equal to the emergent flux.  Above this, the
radius increases rapidly with increasing irradiation.
Expected values for
the maximum incident flux on secondaries in dwarf novae with periods 
longer than 3 hours 
fall below the emergent flux, except perhaps during outburst (Warner 1995),
suggesting that the effects of irradiation do not invalidate our
application of the mass-radius relation for single stars to dwarf novae
secondaries.  

Considered together, the observational and theoretical evidence
suggests that we might expect the results of the exploration we present in 
the following section to apply to those cataclysmic variables which maintain 
modest mass-transfer rates typical of the dwarf nova subclass.  
Those systems with brighter disks and higher inferred mass transfer
rates may contain secondaries larger than implied by our mass-radius 
relationship.  These will still feel the effects of the mass-radius
relationship, but it will be convolved with the effects of their departure from 
thermal equilibrium.  We will not attempt to account for these effects in
this paper.

\subsection{The Orbital Period Histogram}

   In Figure 8 we show the histogram of orbital periods
for non-magnetic cataclysmic variables with known periods.
This plot is reproduced from Warner (1995).  For further
details see also the tabulation by Ritter \& Kolb (1995).
In order to sustain mass transfer 
most cataclysmic variables must lose
angular momentum, either through the emission of gravitational radiation
(Kraft, Matthews, \& Greenstein 1962) or via some other mechanism, and 
thereby evolve to
shorter orbital period.  Thus, while the distribution of periods at 
which cataclysmic variables are born certainly affects the period
histogram, any attempt to fully explain it must account for period 
evolution. 

   At least one feature of the distribution, the minimum
period,  has been accounted for on this basis.  
Paczynski (1967) and Faulkner (1971) recognized 
that at very low mass, the mass-losing 
secondary must become degenerate, and expand in response to 
mass loss.  This corresponds to a change in the sign of slope of the 
theoretical mass-radius relation.  The effect of this change is
to drive the orbital period back to longer
values.  Paczynski \& Sienkiewicz (1983) and Rappaport et al.
(1982) have examined the dependence of the minimum orbital 
period on a variety of physical parameters.  These are important
explorations, because they connect the observations of cataclysmic
variables to the internal physics and composition of the stellar 
models in a fundamental way.  

  Attempts to explain the other most obvious feature, the shortage of 
systems with orbital periods between 2 and 3 hours, are more problematic.
Most explanations derive from a suggestion of Robinson et al. (1981),
who proposed that some mechanism causes secondaries to shrink below their
Roche lobes at a mass of 0.3 $M_\odot$, stopping the mass transfer until
shrinkage of the orbit re-establishes 
it at shorter period. In the interim, the binaries do not appear 
as cataclysmic variables, and the gap is explained.  They 
noted that 0.3 M$_\odot$
is the mass below which models are fully convective and 
speculated that the onset of convection 
in the core might induce the secondary to shrink.
D'Antona \&  Mazzitelli (1982) found a theoretical basis to expect
such shrinkage in the sudden mixing of $^3$He when the star becomes fully 
convective, but later calculations by McDermott and Taam (1989)
showed this effect to be negligible. 

  Most recent models for explaining the period gap
(Rappaport, Verbunt, \& Joss 1983; Spruit \& Ritter 1983; Hameury et al. 1988;
 McDermott \& Taam 1989)
invoke an {\sl ad hoc} ``disruption'' in the angular momentum
loss,  which temporarily separates the secondary from its Roche lobe.  
Because the main angular momentum loss mechanism is believed to be magnetic braking, 
this  disruption requires a decrease  in either the stellar wind or the
magnetic field of the secondary, presumed to be associated with
the onset of convection in the core of the secondary.  
While there are plausibility arguments for 
a disruption of this sort in the magnetic braking
(see Taam \& Spruit 1989), there is no
theory which predicts them, and no convincing observational evidence
that they occur.  On the contrary, observations of low-mass stars
in the Pleiades and Hyades by Jones, Fischer, \& Stauffer (1996)
show no evidence for the reduction of braking in stars below 0.3 $M_\odot$.
For these reasons, the disrupted braking models are somewhat unsatisfying.  
They also preclude applying the information contained in observations
of the period gap to improve structural
models of low-mass stars;  the measurements are used up
in constraining free parameters of the disrupted braking model.

  The disrupted braking models
all require that secondaries of systems with periods just longer than
3 hours be out of thermal equilibrium due to mass transfer.
This is so that a disruption in the mass transfer  at an orbital
period of 3 hours will 
cause the secondary to shrink within its Roche lobe, and remain there
until the orbital period evolves through the gap.  
As we have discussed, it is not clear whether 
all cataclysmic variables meet this requirement. 
 
  The mass-radius relationship we have presented offers the possibility 
to improve upon the disrupted braking models.  We will explore this possibility 
in the next section, first by showing that the features in the mass-radius 
relation correspond to periods associated with the period gap,
and then describing the simple way in which they affect the
orbital period distribution of cataclysmic variables.

\subsection{Orbital Periods and the Mass-Radius Relation}

 The motivation for this investigation derives from a match  
we noted 
between the $M_V$ of the main sequence feature fitted by Reid \& Gizis
(1997), and the $M_V$ of cataclysmic variable secondaries at the 
long end of the period gap as quantified by Warner (1995).
Warner (1995) presented a plot of the absolute magnitudes of
cataclysmic variable secondaries versus  the logarithm of
their orbital periods (his Figure 2.56).
From a linear fit to the data he derived:
\begin{equation}
M_V = 16.7 -1.1~{\rm log} P_{orb}(\rm h).
\end{equation}
Using this equation, the secondary of a cataclysmic variable with
a 3 hour orbit should be $M_V = 11.4$.  From the fit of
Reid \& Gizis (1997) to the 8 parsec data (equation 1), 
the downturn in the color-magnitude diagram starts
at $M_V = 11.6$.   

This correspondence in $M_V$ between the step in the main sequence and the 
period gap suggests that we test for a direct connection in the 
mass-radius plane.
To get orbital periods for a secondary of given mass and radius requires
the use of a formula for the volume-radius of the Roche lobe. 
From Eggleton (1983), we have:
\begin{equation}
{R_L \over a} = {0.49 q^{2/3} \over 0.6 q^{2/3} + \rm{ln}(1+q^{1/3})},
\end{equation}
where $a$ is the orbital separation, and $q$ is the ratio of the primary to the
secondary mass ($M_2/M_1$).  We can apply our measured mass-radius relation
to cataclysmic secondaries through equation 8 by requiring that the radius of 
the secondary $R_2$ be fixed by its mass, $M_2$, according to the mass-radius
relation we have measured.  Under the condition for mass transfer, $R_L = R_2$,
equation 8 then yields the orbital separation.  Using this,
along with Kepler's third law,
\begin{equation}
P_{orb} = \left[{4 \pi^2 a^3 \over G(M_1 + M_2)}\right]^{1/2},
\end{equation}
we get the orbital period.  The 
result is insensitive to $q$ as long as $q < 1$.  We have used a constant
mass for the primary, $M_1 = 0.8
 M_\odot$, in the examples which follow.

Figure 9 shows the mass-radius relation of the Reid \& Gizis (1997) fit to
the 8 parsec sample we calculated in section 2.
For comparison, we have included mass-radius relations which result from 
using theoretical mass-magnitude relations of Kroupa et al. (1993) and
Chabrier et al. (1996) along with the 
empirical one of
Henry \& McCarthy (1993).  
When applied 
to cataclysmic variables, these relations become evolutionary tracks.
The mass-losing secondaries will evolve down the mass-radius
relation until their structure starts to diverge from that
of single stars.  We can calculate their period at each point
using equations 8 and 9.  Furthermore, equations 8 and 9 together
yield the well-known mean density relationship:
\begin{equation}
P_{orb}(\rm h) \approx 8.75 \left({M \over R^3}\right)^{-1/2}
\end{equation} 
(Faulkner, Flannery and Warner 1972; Eggleton 1983),
where M and R are now in solar units. 

 Using equation 10, we have plotted lines of equal orbital period
on Figure 9 to illustrate the effect of changes in the
slope of the mass-radius relationship.  Between the
points labeled 1 and 2, the slow shrinkage of the secondary 
in response to mass loss causes the orbital period to decrease
very slowly.  Between 2 and 3, the orbital period changes more
rapidly for the same amount of mass loss.  This change in the
rate of orbital period evolution generates the period gap in our
model.  

 We can quantify the mechanism illustrated
in Figure 9 by introducing the formula for total orbital angular momentum,
\begin{equation}
J = {M_1M_2 \over M_1 + M_2} a^2 {2\pi \over P_{orb}},
\end{equation}
and assuming that angular momentum is lost from the binary at a constant
rate, $\dot{J}=c$. This assumption is certainly wrong, but will serve
to illustrate how the period gap arises.  
The time spent between points 1 and 2 is given by:
\begin{equation}
t_{1,2}={J_2-J_1\over\dot{J}},
\end{equation}
and likewise for the time between points 2 and 3.
Inserting the appropriate values for $P_{orb}$ (3.47, 3.41 and 2.02 h)
and $J$ (2.35, 1.54, and 0.68 $\times 10^{51}$ g-cm$^2$/s), 
the ratio of the times $t_{1,2}/t_{2,3} = 0.94$, independent of
$c$.  Thus a binary containing a lobe-filling
secondary that obeys the mass-radius relation in Figure 9 
takes almost as long to evolve from 3.47 to 3.41 h as it does 
from 3.41 to 2.02.
This has important implications for the orbital period distribution
of cataclysmic variables.  Under the simplest assumption of
constant cataclysmic variable birth rate, there should be 22
times as many binaries in the bin between 3.47 and 3.41 h
as there are in equal size bins at shorter period (above 2.02 hours).

It is also important to consider what happens to the rate of mass 
loss from the secondary under these assumptions.  
We can calculate this using:
\begin{equation}
\dot{M}_{2;1,2} = {M_{2;2}- M_{2;1}\over t_{1,2}}
\end{equation}
Where $M_2$ refers to the secondary, and the later subscripts refer
to the labeled points in Figure 9.
We find that the mass loss rate for the secondary is slightly higher between 
1 and 2 than it is between 2 and 3 ($\dot{M}_{2;1,2}/\dot{M}_{2;2,3}= 1.13$).
So not only does the binary spend less time between 2 and 3, the mass
transfer rate there is also lower.
Since either effect might contribute to a decrease in the observed number of
cataclysmic variables between  $P_{orb}= 3.41$ and 2.02 h, the change in the 
slope of the mass-radius relation at 0.3 $M_\odot$ provides a general
qualitative reason to expect a reduction in the observed numbers 
of cataclysmic variables between 3.41 and 2.02 h.  
In the next section, we will introduce the
more realistic assumption that angular momentum loss
decreases 
slowly with time, as enforced by a prescription for magnetic braking.

\subsection{An Illustration Using Magnetic Braking}

   The contents of this section are meant as an illustration only.  Many of
the results are highly sensitive to the exact slope of the measured 
mass-radius relation, which is not well-constrained due to observational noise.
Nonetheless, the results 
illustrate the general shape of the orbital period distribution and mass loss
rates implied by the mass-radius relation, using more sophisticated
assumptions than in the previous section.
If the model for the period gap we present
survives continued scrutiny, the sensitivity of the
the results in this section to the mass-radius relation will
eventually be beneficial. It will allow even crude measurements
of cataclysmic variable properties to constrain
low-mass stellar models.

  Our calculations in this section are based 
on the formulae of the previous section, but with a finer
spacing in mass.  From these formulae we have calculated sample 
orbital period distributions and mass transfer rates 
using a parameterized angular momentum loss
rule of the form,
\begin{equation}
\dot{J}= -3.8 \times 10^{-30}~MR_\odot^4 \left({R \over R_\odot}\right)^\gamma\left({2\pi \over P_{orb}}\right)^3~{\rm dyn-cm},
\end{equation}
following Rappaport et al. (1983), who rely upon Verbunt \& Zwaan 
(1981).  We do {\it not} introduce artificial reductions in this loss,
as required for most models of the period gap.  Consequently, our models
include angular momentum loss in excess of the rate caused by gravitational
radiation, even below the period gap.  This is in better agreement with the
observations of accretion disk magnitudes below the period gap, which 
frequently require mass transfer rates too high to be driven by 
gravitational radiation alone (Warner 1995).

  Figure 10 shows the results of calculations using the mass-radius
relation we constructed from the Reid \& Gizis (1997) fit to the 8 parsec data,
and the empirical (mass, $M_V$) relation of Henry \& McCarthy (1993).
This relation is the dotted line in Figures 3, 4 and 9.  We used the
angular momentum loss parameterization of equation 14 with $\gamma=4$
and included 
gravitational radiation loss according to Landau \& Lifschitz (1958).
We fixed the mass of the white dwarf primary at 0.8 $M_\odot$, equivalent to
assuming all the transferred mass is eventually lost from the system.
Calculations using conservative transfer yield almost identical results.

 The top panel of Figure 10 shows the time spent at each orbital period,
from equation 12.  These times have been normalized to equal orbital period
intervals to allow direct comparison to the orbital period histogram (bottom 
panel). Under the assumption of uniform cataclysmic variable birthrate (with initial 
periods above 5 hours), this curve 
represents the space density of cataclysmic variables at each orbital period.
We calculated curves using the other two mass-radius relations in 
Figure 9, with similar results.  The curve based on the Kroupa et al.
(1993) (mass, $M_V$) relation gives a broader and shorter peak at 3.5 h,
but still has small values at longer period, while the Chabrier et al.
(1996) based relation yields a larger peak at 3.5 h.  
If there were no selection effects involved in the discovery of cataclysmic
variables, these
curves would not be promising;  the 
bump near 3.5 h is too narrow, and values for longer period are too low
to match the observed orbital period histogram.

 However, the middle panel of Figure 10 shows the secondary mass loss rate 
as a function of orbital period, from equation 13.  The increase in the
mass loss near 3.5 h is a real effect of the change in slope of
the mass-radius relation.  The subsequent drop is no more rapid
than before, but is compressed on this plot because of the long time spent 
by the model in the vicinity of $P_{orb} = 3.5$ h.
Because higher mass transfer rates result in  
brighter disks, and increase the likelihood of detection,
the middle curve in Figure 10 represents an important
factor in the discovery probability for cataclysmic variables.
Obviously, the higher mass transfer rates at long period
can compensate, to some degree, for the shortage in our
model space densities in the same portion of the upper panel.
The degree of compensation is impossible to calculate without 
precise knowledge of cataclysmic variable discovery selection
effects.  Considered together, the panels in Figure 10 suggest
that the mass-radius relation we have constructed 
can account for the main features of the cataclysmic variable 
orbital period distribution.

\section{Summary and Conclusions}

 We have shown that the color-magnitude diagram of the
lower main sequence exhibits a feature which is apparently
associated with changes in the slope  of the mass-radius relation for
low-mass stars.  Investigations in the mass-radius plane show that the
locations of the interesting features correspond roughly to the 
boundaries of the known cataclysmic variable orbital 
period gap. This correspondence
relies upon the best available empirical data:
the observed color magnitude of the 8 parsec sample from
Reid \& Gizis (1997), the bolometric corrections of Leggett (1995),
the mass-visual magnitude relation of Henry \& McCarthy, and the
measured orbital periods of cataclysmic variables (Warner 1995; Ritter \& Kolb 1995).

  By introducing additional theoretical assumptions, chiefly regarding the
structural similarity of low-mass single stars and cataclysmic variable 
secondaries, we find that the observed mass-radius relation can  
generate a
period gap similar to the one observed.
Unlike disrupted braking models,
this explanation does not require any {\sl ad hoc} adjustments
to the angular momentum loss from the binary.  It also does not demand that
all secondaries above the period gap be out of thermal equilibrium, a 
requirement of disrupted braking models that is in conflict with the 
observations. Furthermore, the explanation we have uncovered allows
angular momentum loss below the period gap in excess of that caused by 
gravitational radiation, in better agreement with the observations
(Warner 1995).

  Our explanation does not apply directly to systems with high mass transfer
rates, which may have secondaries larger than single stars of equivalent mass.
These systems will still be affected by changes in the mass-radius relation, 
but in ways that are complicated by the fact that the secondaries are not 
necessarily in thermal equilibrium.  We have not attempted to unravel these 
effects in this paper.

  Our investigation suggests two further lines of inquiry, the first of which
is completely theoretical.  Given the uncertainties in the transformations
we have applied to the data, it is possible that
the changes in the mass-radius relation are artifactual.  
If it is possible to show theoretically that slope changes such as those 
we observe in the mass-radius
relation are disallowed, we will be forced to abandon any further
pursuit of their relationship to cataclysmic variables 
and attribute the connection explored in this
paper to an appealing, but accidental, coincidence.
However, theoretical investigations of this sort would have 
another important use, arising from constraints they would  
provide
on the combined color-temperature and mass-magnitude relations.  These might
improve the reliability of investigations of the initial mass function and the 
luminosity function of low-mass stars.

  The second course to pursue is observational: primarily, to improve the
measurements of the mass-radius relationship.  There is a notable
shortage of fundamental mass-radius measurements, acquired from 
eclipsing binaries where the masses and radii are both directly
measurable.
The only system available below 0.5 $M_\odot$ is CM Dra, and there is evidence
that this system may be metal poor (Metcalfe et al. 1996; Gizis 1997).  
It is possible that the 
eclipsing cataclysmic variables themselves may contribute to
this effort. If independent arguments can establish that cataclysmic
variable secondaries
follow a normal main sequence mass-radius relationship, the 
data from Webbink presented in Figure 7 offer an order of magnitude increase 
in the number of direct empirical mass-radius determinations. 

  Finally, if the results of this paper are correct, there are 
extensive numerical experiments to conduct. Hameury (1991) has emphasized 
the value of cataclysmic variables for testing low-mass stellar models;
cataclysmic variable properties are highly sensitive to the physics governing
the secondary stars.  If future investigations can use the information
provided by the size and location of the period gap, without 
introducing {\sl ad hoc} components to the theory, it  
will be a valuable addition to the
very few measurables offered by single stars. Since much of the physics
governing stellar interiors is inaccessible by any other means, 
it is important to explore every observational constraint our ingenuity can 
devise.

\section{Acknowledgments}

  We are indebted to the referee R. Webbink, who provided us his data on
masses and radii of cataclysmic variable secondaries. His comments 
resulted in substantial improvements to this paper, both in content and 
presentation. 

  This work was begun with support from NASA through grant number HF-01041.01-93A from the Space Telescope
Science Institute, which is operated by the Association of Universities for Research in Astronomy, Inc.,
under NASA contract NAS5-26555; it was completed with the generous support of the Sherman Fairchild 
Foundation.  INR acknowledges partial support from NASA through grant GO-05913.01-94A.
JEG gratefully acknowledges partial support from both a Greenstein Fellowship 
and a Kingsley Fellowship. MSO'B receives support from a GAANN fellowship, through grant
number P200A10522 from the Department of Education.

\clearpage
\centerline{FIGURE CAPTIONS}
\vskip2em

%figure 1
\figcaption{Color magnitude diagrams for stars in the 8 parsec sample
of Reid \& Gizis (1997).  Open circles indicate data
expected to have larger photometric errors (see text). The lower 
panel shows the same data
as in the upper left, but
incorporating better distances from Hipparcos, where available.}

%figure 2
\figcaption{Comparison between the observations and theoretical models.
Both panels show 
$M_V$, ($V-I$) measurements from the 8 parsec sample,
transformed into log($L/L_\odot)$ using data from Leggett et al. (1996) ({\sl filled
and open circles}),
and the fit of Reid \& Gizis (1997) ({\sl dotted line}). Theoretical models are 
divided between the plots as follows: {\sl upper squares}: Dorman et al. (1989)
with the Fontaine, Graboske, \& Van Horn (1977) equation-of-state; 
{\sl upper triangles}: D'Antona \& Mazzitelli (1994);
{\sl upper pentagons}: Neece (1984); {\sl lower squares}: Baraffe \& Chabrier 
(1996); {\sl lower triangles}: Baraffe et al. (1995). }

%figure3
\figcaption{Mass-radius relation for the $M_V$, ($V-I$) 8 parsec data 
({\sl open and filled
circles}) and for the models of Baraffe \& Chabrier (1996)({\sl solid line}).
Masses are from the (mass, $M_V$) relation of Henry \& McCarthy (1993).
The {\sl dotted line} is the fit of Reid \& Gizis (1997) transformed in the
same manner as the data.  See the text for a discussion of the error bars.}

%figure4

\figcaption{Mass-radius relation of the $M_K$, ($I-K$) 8 parsec data
({\sl open and filled circles}), and the data tabulated by Popper (1980)
({\sl filled squares}).  For reference, the $M_V$, ($V-I$) fit of
Reid \& Gizis (1997) ({\sl dotted line}) remains as in Figure 4.
The eclipsing, spectroscopic binaries YY Gem and CM Dra are marked.
For the latter we have plotted the newer masses and radii from Metcalfe
et al. (1996)}

%figure5
\figcaption{The color-temperature transformation ({\sl filled circles}) that would be required to
bring the observational data into exact correspondence with the models of Baraffe \& 
Chabrier (1996).  The transformations of Bessell (1995)({\sl solid line})
and our fit to Leggett et al. (1996)({\sl dotted line}) 
are shown for reference.}

%figure6

\figcaption{Mass-radius relations for a variety of models.
They are: {\sl solid line}: Baraffe \& Chabrier (1996); {\sl dotted line}:
Tout et al. (1996); {\sl short dashes}:
D'Antona \& Mazitelli (1994); {\sl long dashes}: Dorman et al. (1984); 
{\sl dot-short dashes}: Neece (1984)}

%figure7

\figcaption{The mass-radius relation of eclipsing cataclysmic variables
from Webbink (1990; 1997)({\sl large circles}).  For comparison, the
{\sl small circles} are the mass-radius relation for the 8 parsec sample
shown previously in Figure 3.  The fit of Reid \& Gizis 
(1997)({\sl dotted line}) 
and the models of Baraffe \& Chabrier (1996)({\sl solid line}) are also plotted
as in Figure 3.}

%figure8

\figcaption{The orbital period distribution of non-magnetic
cataclysmic variables.  Adapted from Warner (1995).}

%figure9

\figcaption{The mass-radius relation and its application to
cataclysmic variable secondaries.  The curves are the transformed
fit of Reid \& Gizis (1997) with masses from the (mass, $M_V$)
relations of: {\sl dotted line}: Henry \& McCarthy (1993);
{\sl dashed line}: Kroupa et al. (1993); and {\sl dot-dash}:
Chabrier et al. (1996). The solid lines are lines of constant orbital 
period from equation 10. See the text for  discussion of the points
labelled 1, 2 and 3.}

%figure10

\figcaption{Orbital period and mass-loss evolution 
calculated from the observed mass-radius relation. 
{\sl Top Panel}: Time spent between  
orbital periods
normalized by the period interval (see equation 12). {\sl Middle panel}:
secondary mass loss rate (see equation 13). {\sl Bottom panel}:
Orbital period histogram for all non-magnetic cataclysmic variables
({\sl solid line}), and for dwarf novae only ({\sl filled regions}).}

\clearpage

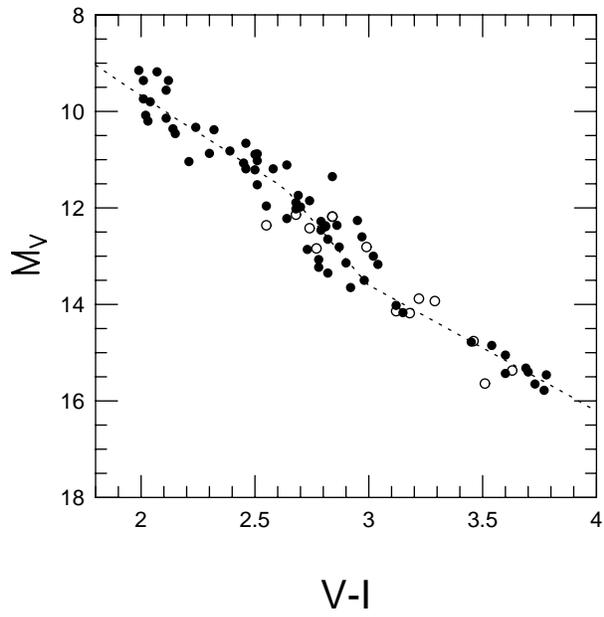
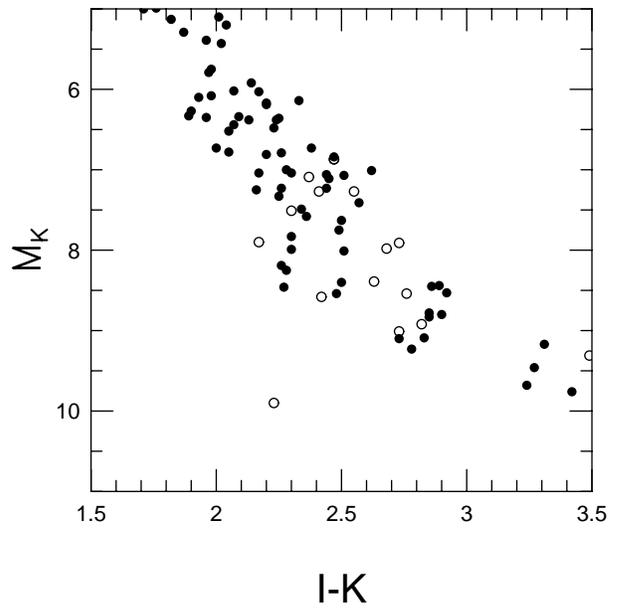
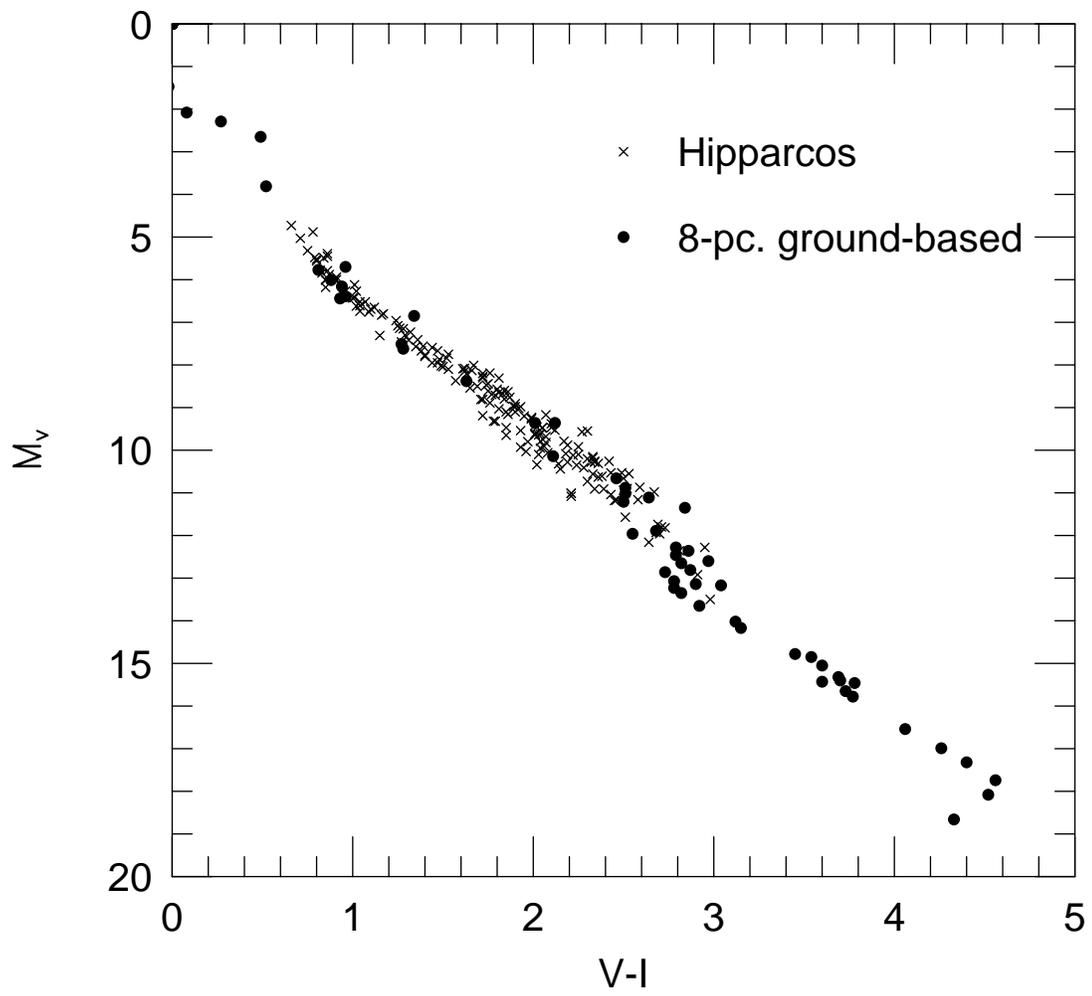
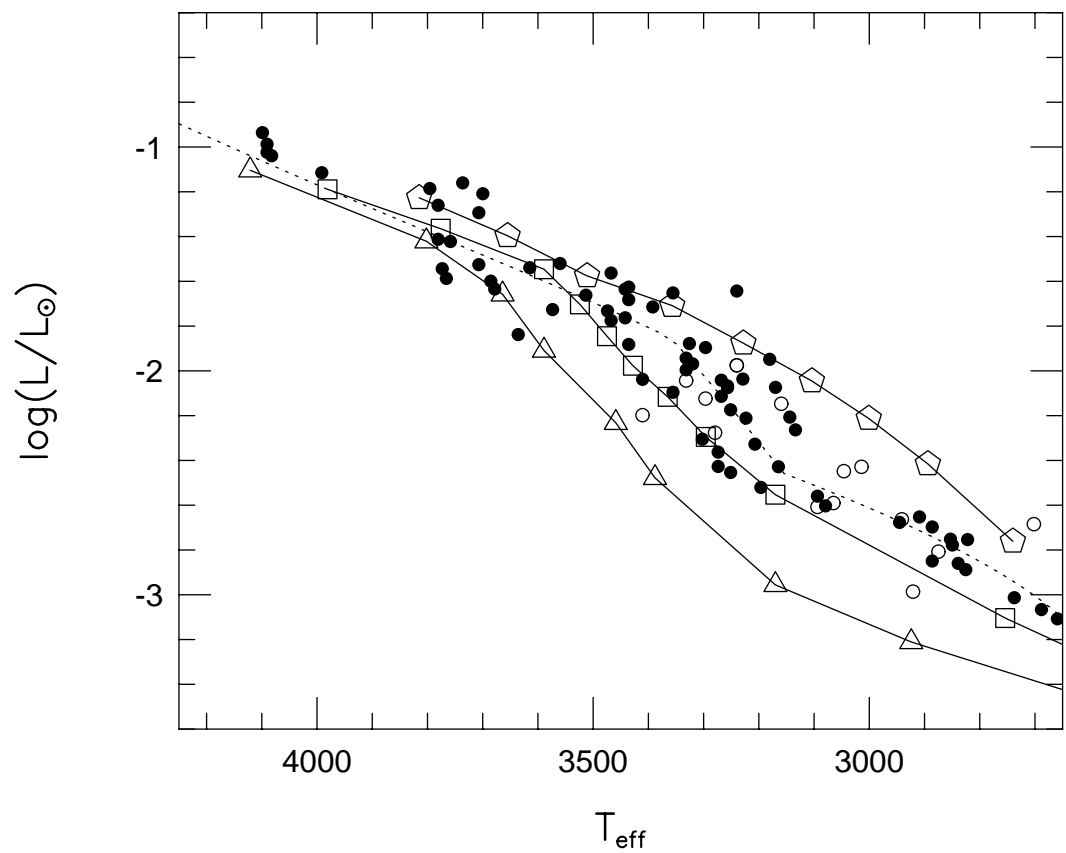
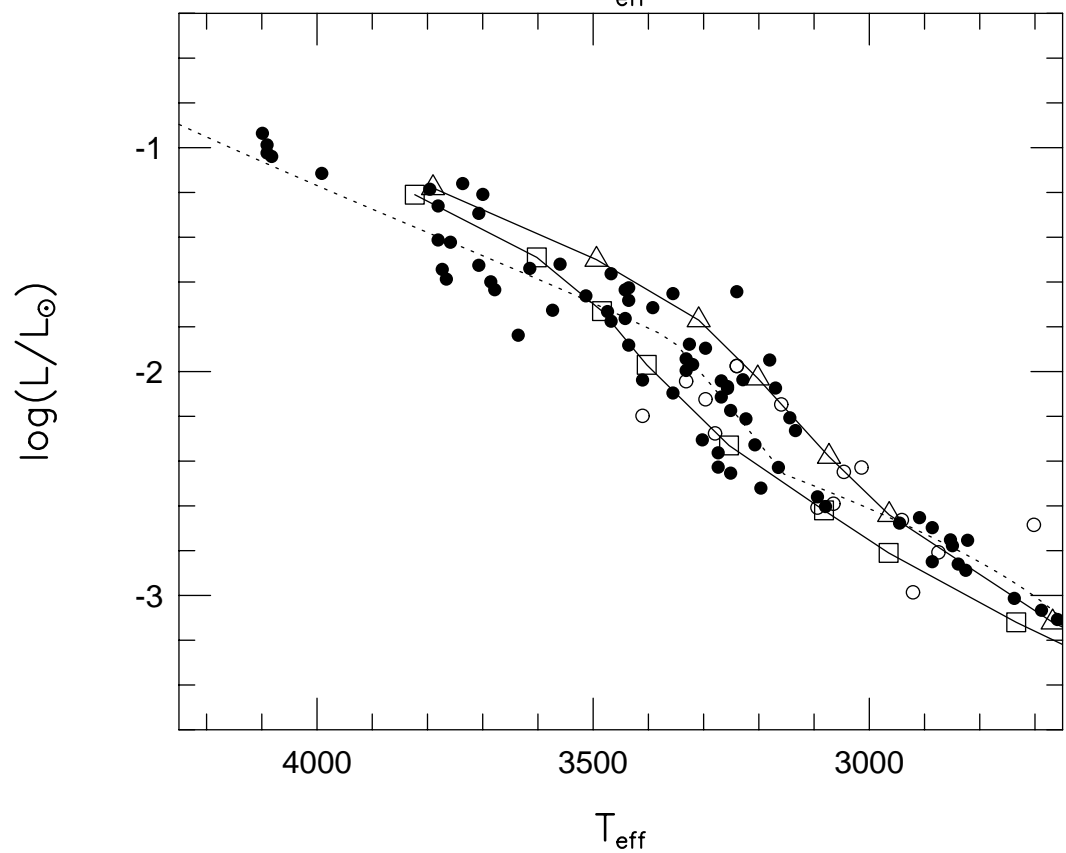
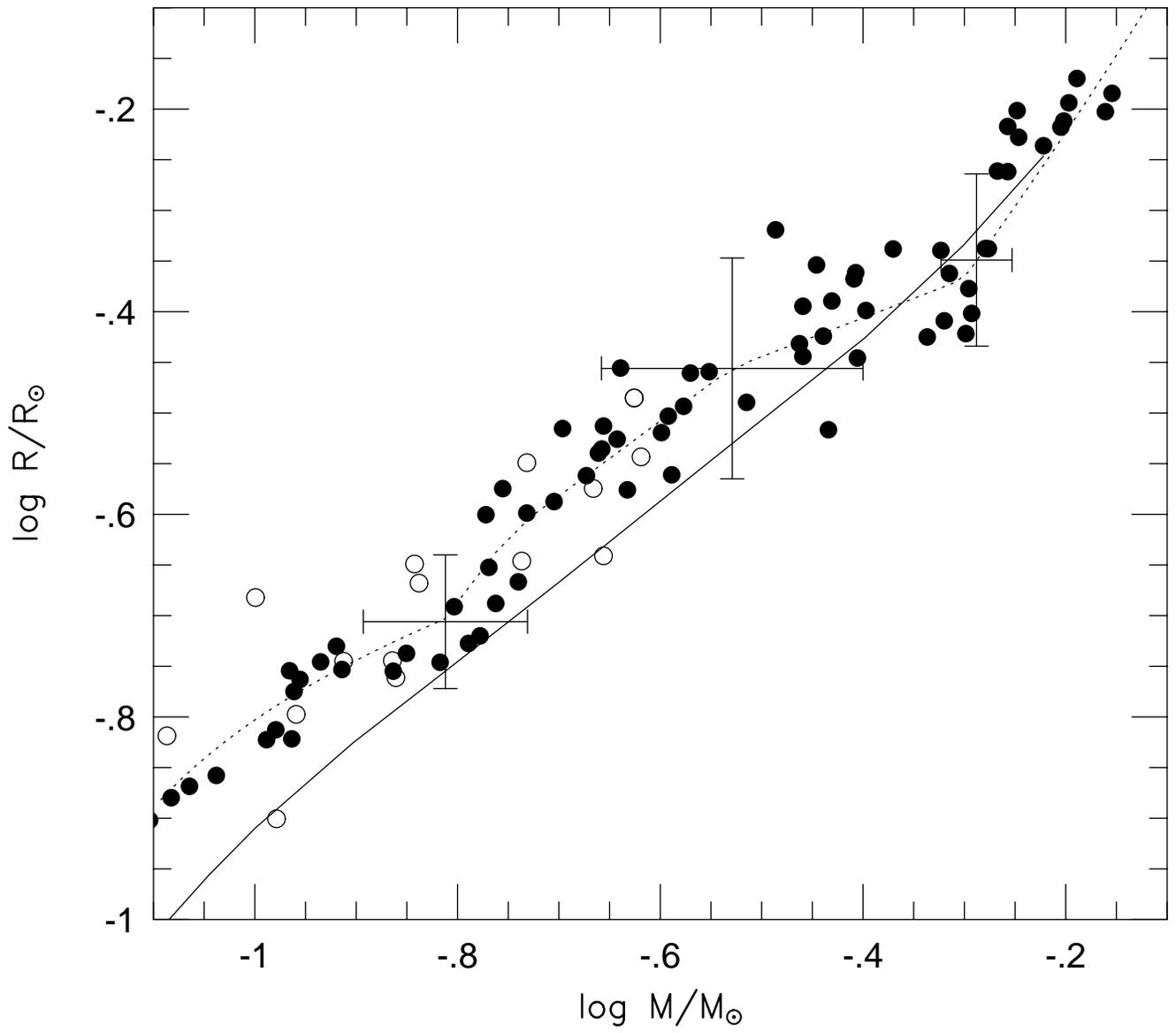
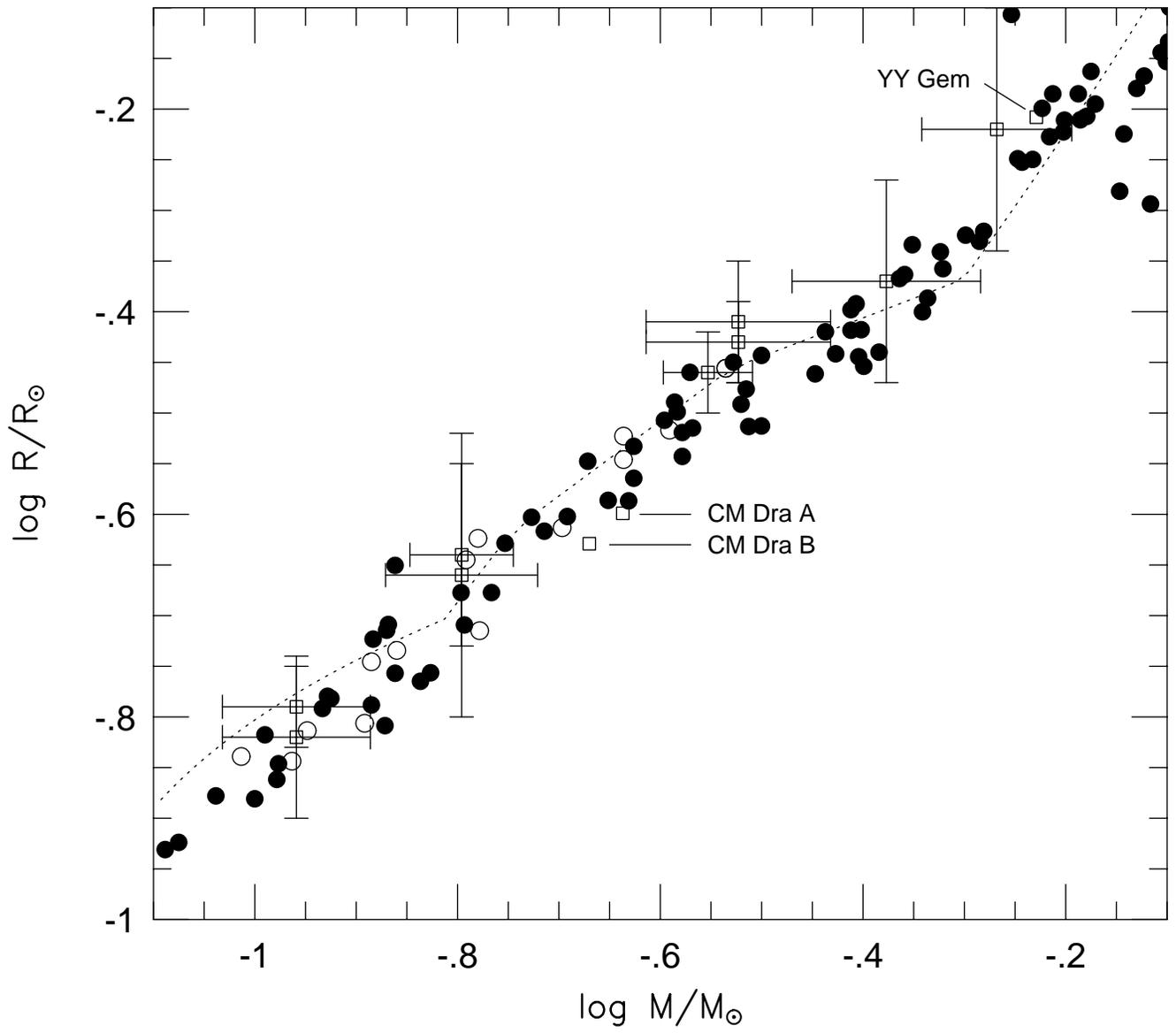
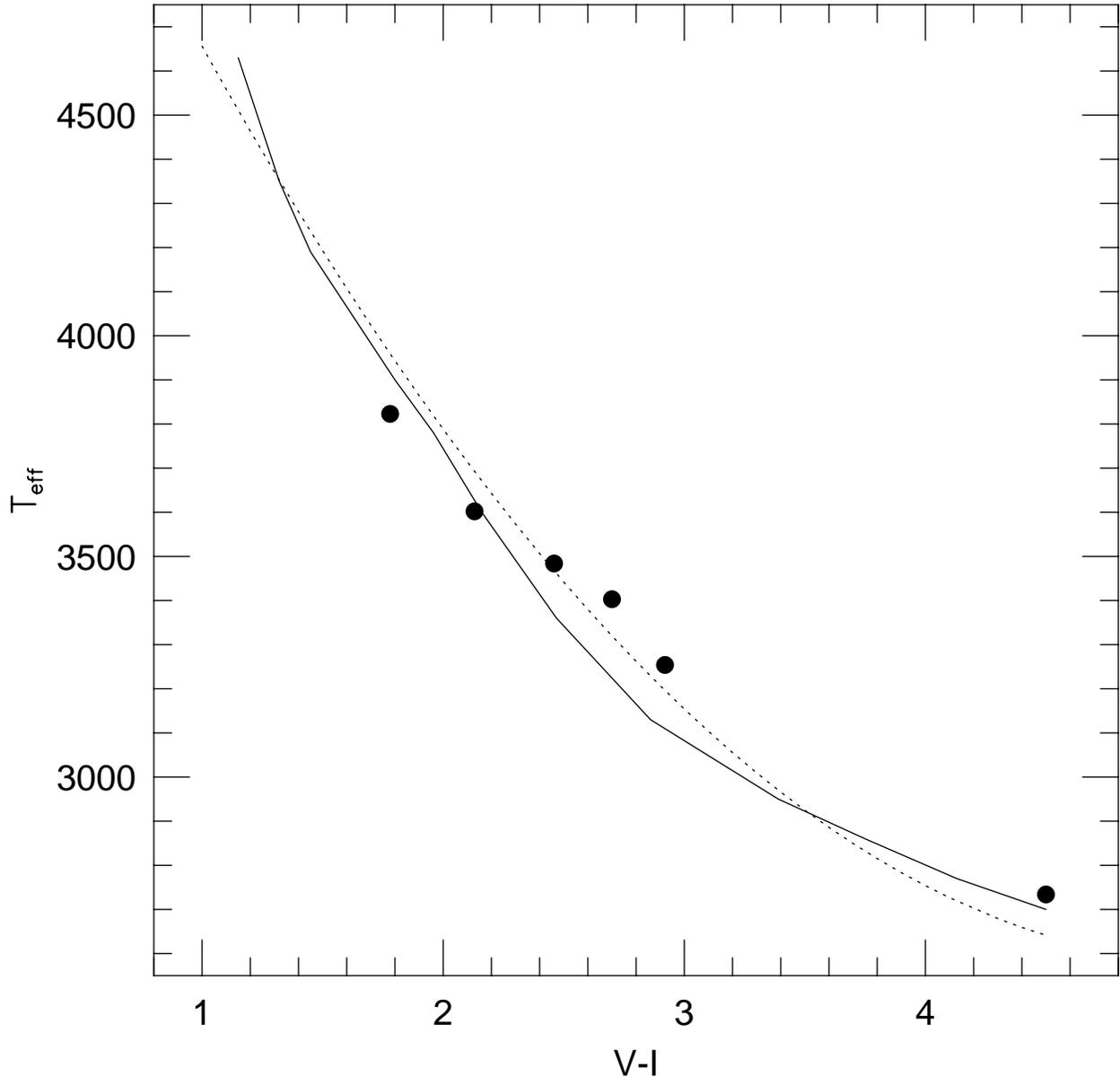
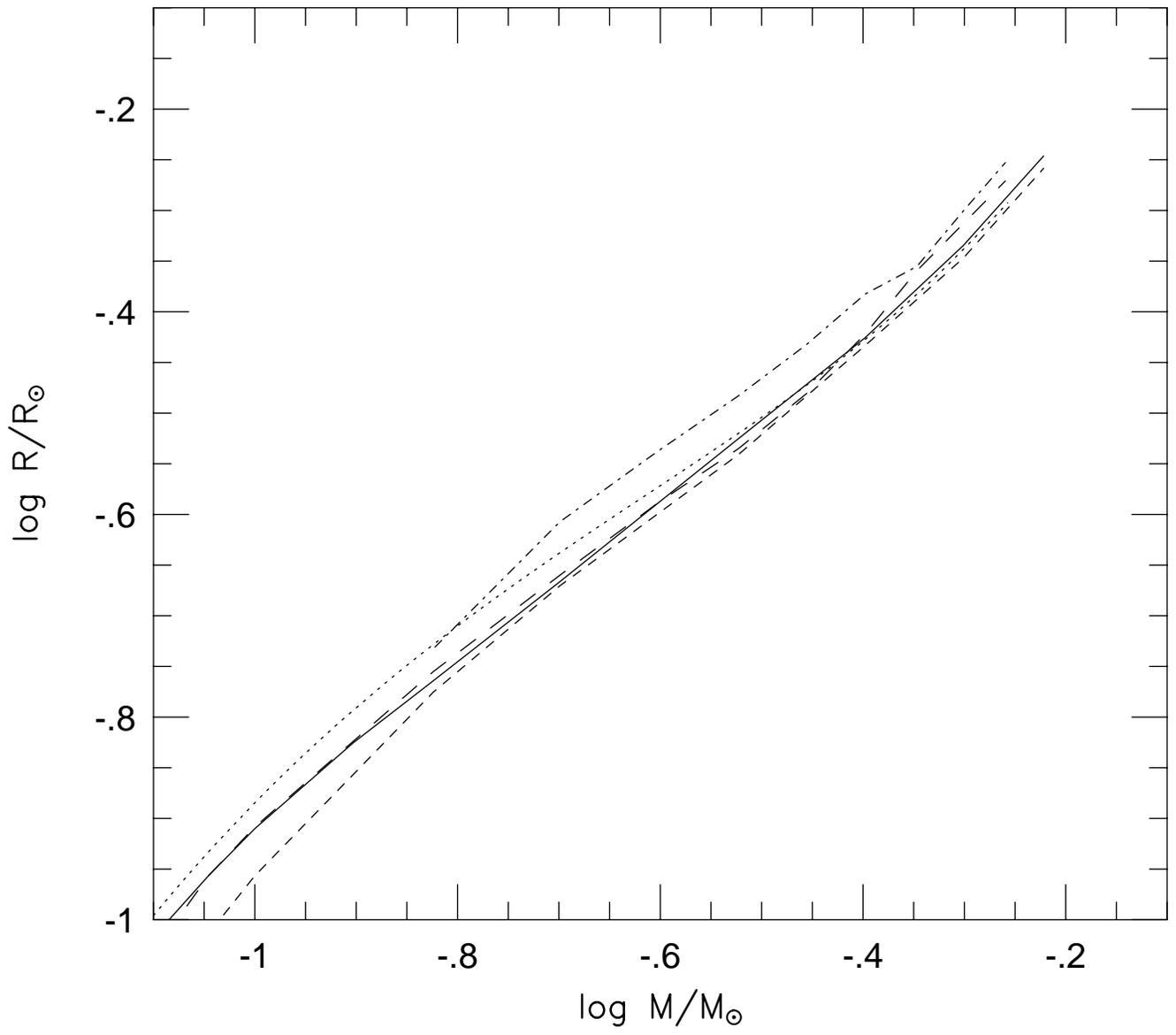
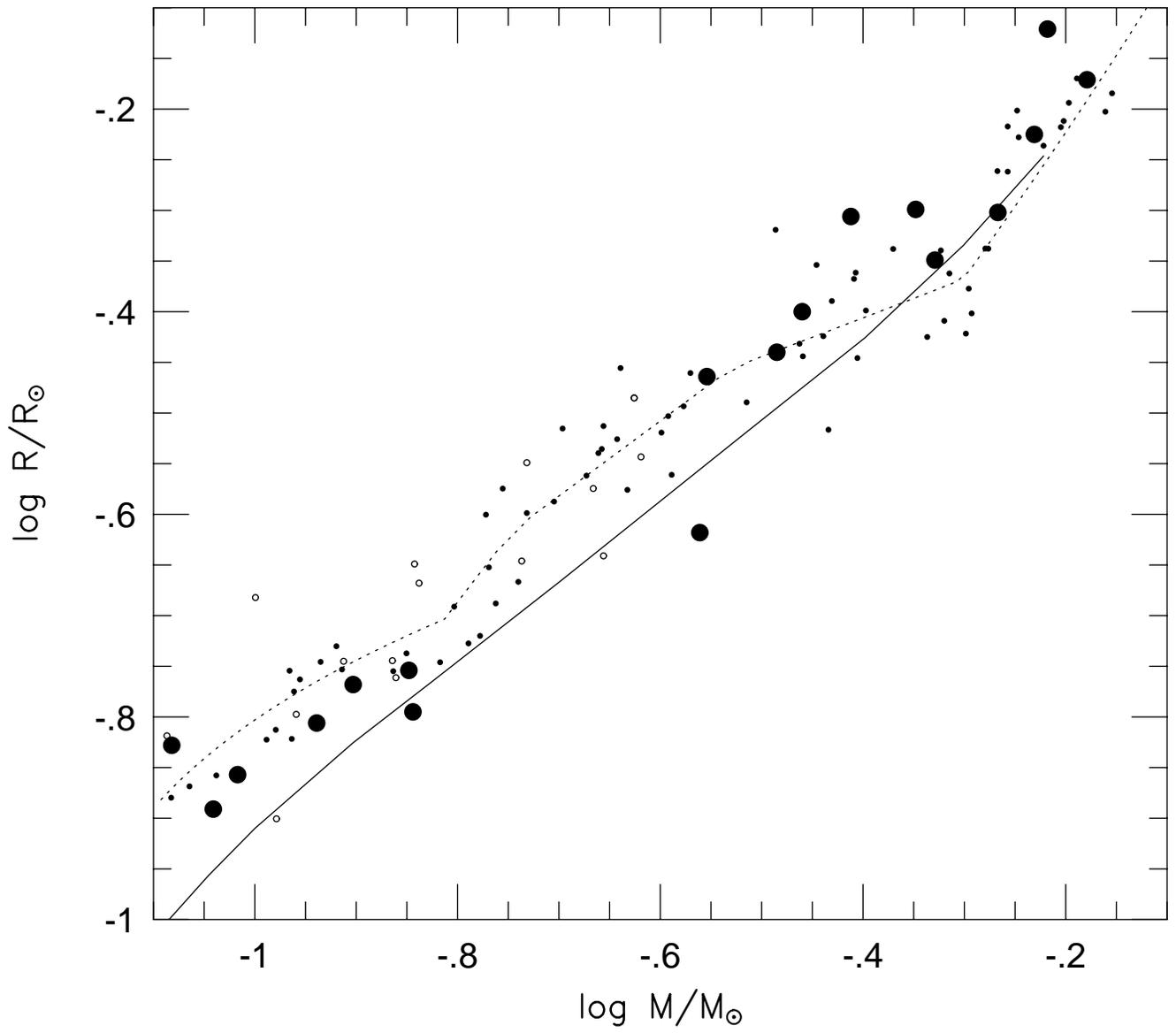
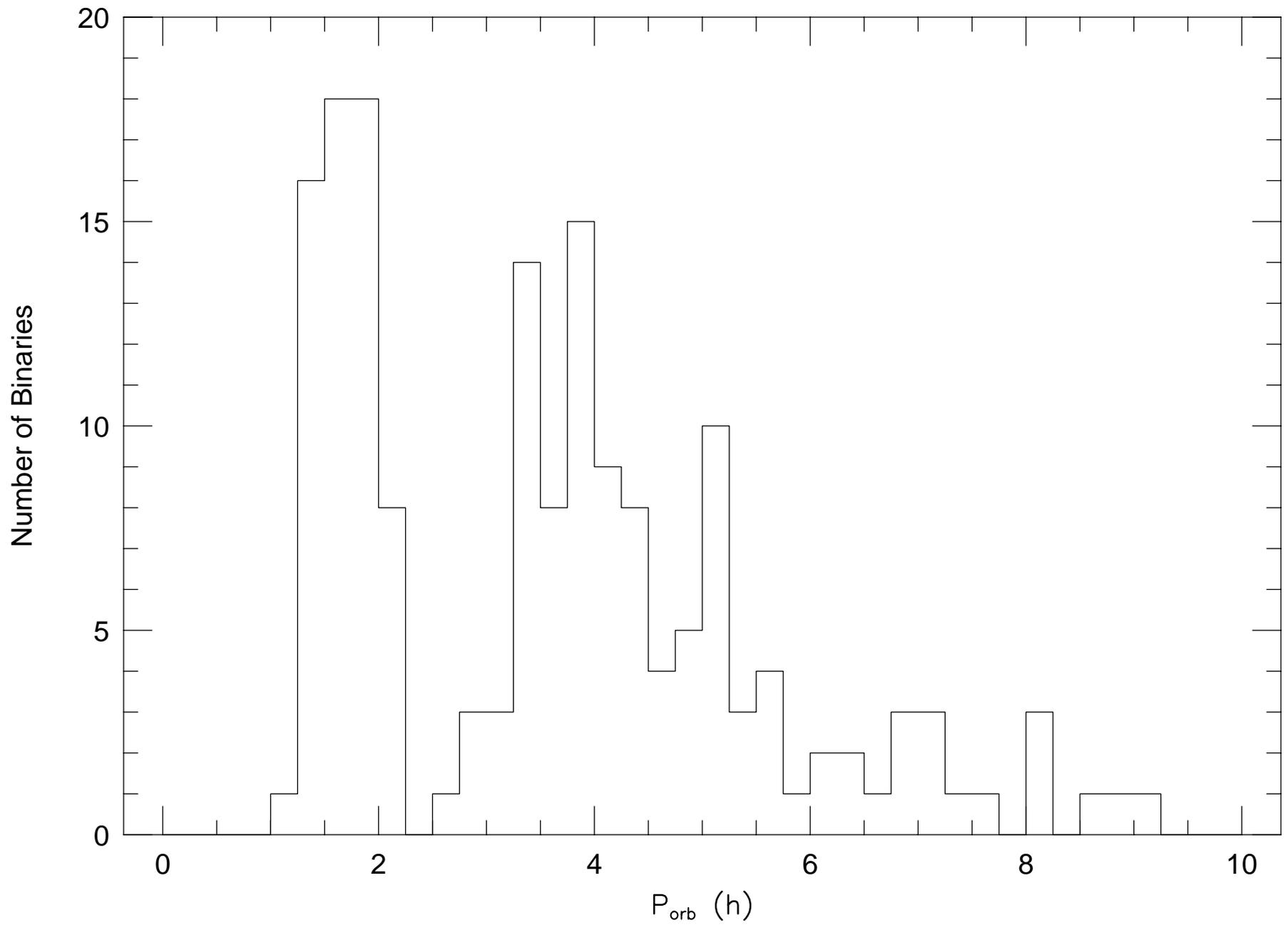
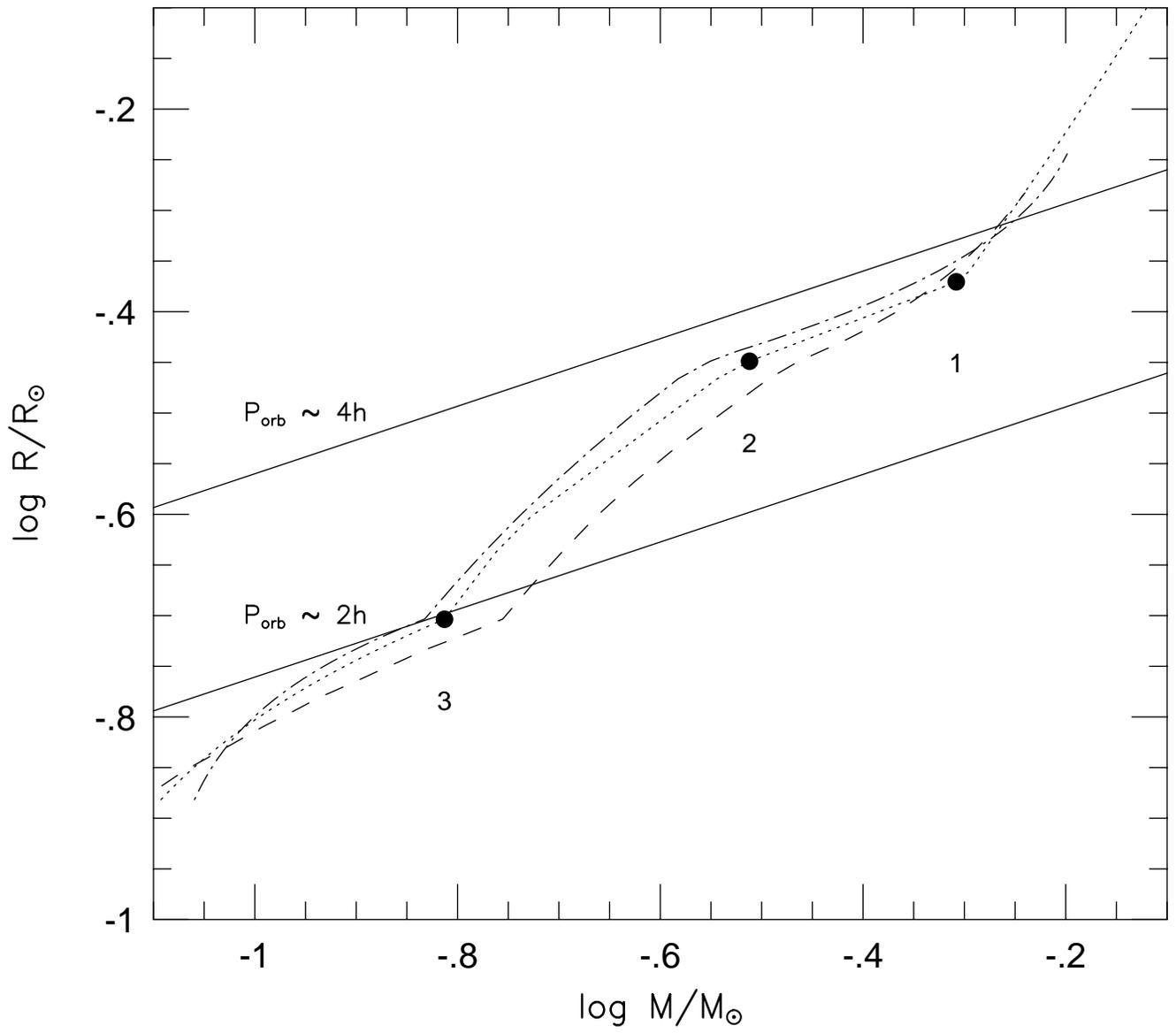
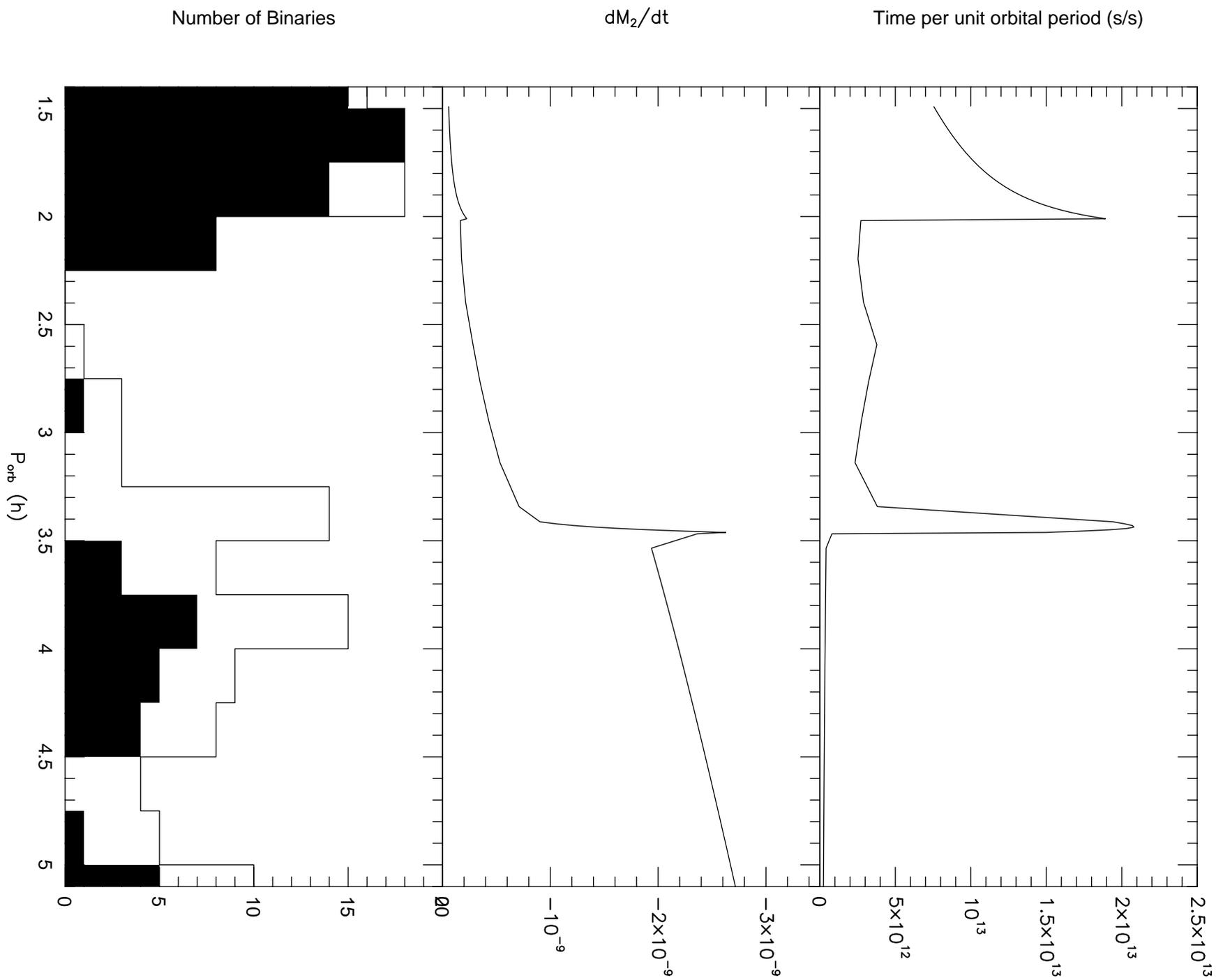
\end{document}